\documentclass[conference]{IEEEtran}
\IEEEoverridecommandlockouts

\usepackage{cite}
\usepackage{amsmath,amssymb,amsfonts}
\usepackage{algorithmic}
\usepackage{graphicx}
\usepackage{textcomp}
\usepackage{xcolor}
\usepackage{booktabs}

\usepackage[acronym]{glossaries}
\makeglossaries

\def\BibTeX{{\rm B\kern-.05em{\sc i\kern-.025em b}\kern-.08em
    T\kern-.1667em\lower.7ex\hbox{E}\kern-.125emX}}
\begin{document}

\newacronym{rtt}{RTT}{Round-Trip Time}
\newacronym{drl}{DRL}{Deep Reinforcement Learning}

\newacronym{mae}{MAE}{Mean Absolute Error}
\newacronym{mse}{MSE}{Mean Squared Error}
\newacronym{mape}{MAPE}{Mean Absolute Percentage Error}

\newacronym{vcpu}{vCPU}{virtual Central Processing Unit}
\newacronym{ram}{RAM}{Random Access Memory}

\newacronym{mlp}{MLP}{Multi-Layer Perceptron}
\newacronym{mac}{MAC}{Medium Access Control}

\newacronym{veco}{VECO}{Vehicular Edge Caching and Offloading}

\title{Closing the Loop: Continuous Measurement-Driven Refinement of Offloading Predictions\\
}
\author{
\IEEEauthorblockN{Falk Dettinger, Matthias Weiß and Michael Weyrich}
\IEEEauthorblockA{
\textit{Institute of Industrial Automation and Software Engineering (IAS)} \\
\textit{University of Stuttgart} \\
Pfaffenwaldring 47, 70569 Stuttgart, Germany \\
E-Mail: \{falk.dettinger, matthias.weiss, michael.weyrich\}@ias.uni-stuttgart.de}}

\maketitle

\begin{abstract}

Modern vehicles increasingly offload computationally intensive perception and decision functions to backend servers, requiring accurate predictions of 
absolute performance metrics such as \glsentrylong{rtt} (RTT), processing time, and utilization. In practice, strong temporal variability, heterogeneous backend hardware, and 
multimodal latency regimes cause offline-trained predictors to drift, creating a reliability gap for latency-sensitive functions.
We address this gap with an operational, measurement-driven closed loop that continuously recalibrates absolute-value predictors during runtime. The 
system aligns real execution measurements with predicted values and performs incremental online updates of a lightweight multi-head neural network while 
preserving model stability. The model implicitly learns the broad, non-Gaussian spread of input metrics, and a sigma-based error analysis in our 
evaluation characterizes residual variability under dynamic conditions.
Experiments across two Kubernetes clusters show that continuous measurement-driven refinement reduces prediction drift, improves accuracy for RTT, 
processing time, and utilization, and stabilizes prediction behavior across heterogeneous latency regimes. However, the broad and multimodal 
distribution of input metrics imposes fundamental limits on absolute-value prediction, with residual errors frequently exceeding configured 
thresholds. Overall, online calibration proves feasible and necessary for robust computation offloading in dynamic vehicular edge environments, 
while highlighting the need for future mechanisms that address extreme latency regimes and high-variance operating conditions.

\end{abstract}


\section{Introduction}

Modern vehicles increasingly rely on computationally intensive perception, prediction, and decision-making functions \cite{baumann2024total} that exceed the capabilities of static onboard hardware, which remains 
unchanged for more than a decade \cite{SPGlobal2024}. This mismatch threatens the performance of safety-critical functions including advanced driver-assistance systems and real-time perception tasks \cite{Mizrachi2025}. 
Computation offloading to edge or cloud servers offers a promising solution \cite{Stuempfle2025SDV, praveen2025autonomous}, but automated vehicle functions operate under strict real-time constraints and highly dynamic 
connectivity conditions \cite{fi16040108, weiss2025sdvdiag}. Server availability, signal quality, and backend load can change within seconds, making reliable offloading decisions dependent on accurate predictions of 
absolute performance metrics such as \gls{rtt}, processing time, and utilization.

Existing offloading approaches typically rely on offline-trained models \cite{Stuempfle2025SDV}, simplified latency assumptions, or reward-based learning signals. These methods overlook a fundamental challenge: 
in real systems, performance characteristics drift over time due to mobility, contention, and heterogeneous hardware \cite{fi16040108, dettinger2025directives}. As a result, prediction models degrade unless 
they are continuously recalibrated. Prior work has introduced conceptual feedback mechanisms \cite{mohammadi2025adaptive, yang2021peer, adu2026decentralized}, but none has demonstrated a fully implemented, 
measurement-driven closed loop that updates predictive models themselves based on real execution data.

This gap motivates the central research question of this work: \textit{How can a computation offloading pipeline continuously align its predictive models with real system behavior to maintain reliable 
absolute-value predictions under dynamic execution conditions?}

To address this question, we introduce an operational, measurement-driven feedback loop for computation offloading pipelines. Our system continuously collects execution measurements from heterogeneous 
backend servers, aligns them with predicted values, and incrementally updates a lightweight multi-head \glossary{mlp} during runtime. This closes the missing link between the Prediction Layer and the actual system 
behavior, enabling the pipeline to correct systematic errors, mitigate drift, and adapt to fluctuating network and load conditions without retraining the entire model.

The remainder of this paper is organized as follows. Section~\ref{sec:background} revisits the original pipeline and highlights why online calibration is required. The related work regarding the feedback 
integration in computation offloading is provided in Section~\ref{sec:related_work}. We then formalize the prediction task and the feedback setting in Section~\ref{sec:problem}. The proposed online-learning 
feedback loop is detailed in Section~\ref{sec:design}, followed by an evaluation under real execution conditions in Section~\ref{sec:evaluation}. The paper closes with a discussion of the results in Section 
\ref{sec:discussion} and a conclusion in Section~\ref{sec:conclusion}.

\section{Background and Limitations of Previous Pipeline} \label{sec:background}

In our previous paper \cite{dettinger2026intelligentcomputationoffloadingdynamic}, we presented a four-layer computation offloading pipeline for Software-Defined Vehicles. Its structure 
spans from information aggregation to execution and feedback. At the front end, the Extraction Layer collects function requirements and backend server metadata, forming the basis for all subsequent decisions.

Building on this information, the Decision Layer determines where a function should run. It first predicts key performance metrics, \gls{rtt}, processing time, and utilization, and then 
evaluates feasible servers by combining these estimates with mobility-related factors such as direction, distance, and remaining stay time. This two-stage process enables the system to select an 
appropriate backend target or fall back to local execution when necessary.

Execution itself is handled by the Offloading and Execution Layer, which relies on a Kubernetes-based backend to orchestrate microservices and support both local and remote processing.

The final component, the Detection Layer, compares predicted and measured metrics to assess decision correctness. In the previous pipeline, this mechanism existed only conceptually and was not active 
during evaluation, leaving the system without continuous recalibration. The present work closes this gap by introducing an online-learning feedback loop that updates latency and load predictions during 
runtime and feeds refined estimates back into the Decision Layer.

\section{Related Work} \label{sec:related_work}

Computation offloading in dynamic fog, edge, and vehicular environments has been explored from several complementary perspectives, each relying on different forms of feedback. A first line of research approaches 
offloading as a sequential learning problem. Bandit-based methods \cite{yang2021peer,zhu2019online} exploit delayed or partial observations to refine offloading policies, while reinforcement-learning techniques 
\cite{mohammadi2025adaptive} adapt decisions through reward signals. More recent work integrates digital twins into \gls{drl} pipelines \cite{liang2025digital} to generate synthetic predictive feedback, and multi-agent 
\gls{drl} frameworks \cite{adu2026decentralized} incorporate \gls{mac}-level latency dynamics directly into the learning process. Hybrid-action \gls{drl} \cite{zhang2025e2e} further combines discrete offloading choices with continuous 
resource control. Although these approaches make extensive use of feedback, it is primarily used to optimize policies rather than to recalibrate predictive system models.

A second research direction focuses on mobility-aware and dynamic vehicular edge computing. Structural feedback has been used to guide multi-objective optimization of offloading and caching \cite{qiu2025joint}, while 
predictive trajectory estimation \cite{geng2025computation} helps mitigate mobility-induced delays. Digital-twin-based systems such as \glossary{veco} \cite{lin2025veco} provide synthetic estimates of network and resource states 
to support adaptive decisions. These mechanisms, however, remain model-driven and do not incorporate real execution measurements to correct latency or utilization predictions.

A third strand examines offloading in federated or trust-sensitive environments, where decentralized coordination and heterogeneous resource availability introduce additional uncertainty. Architectural challenges are 
discussed in \cite{pournazari2025computation}, and trust-aware mechanisms \cite{wang2025trust} use reputational signals to stabilize decisions in the presence of unreliable edge servers. While such approaches improve 
robustness, they do not update predictive performance models based on real-time measurements.
\vspace{6px}

Across all three strands, existing work employs delayed, partial, structural, predictive, synthetic, reputational, or reward-based feedback. What is consistently missing is a mechanism that continuously aligns 
predictive models with real system behavior. This work closes that gap by introducing a lightweight, measurement-driven feedback loop that updates predictive models directly from execution data. To the best of 
our knowledge, no prior work updates absolute-value predictors directly from real execution measurements in a closed loop in the context of computation offloading.

\section{Problem Formulation: Absolute-Value Prediction under Feedback} \label{sec:problem}

Computation offloading in dynamic edge environments requires accurate predictions of absolute performance metrics \cite{dettinger2026intelligentcomputationoffloadingdynamic} such as \gls{rtt}, processing time, and system 
utilization. Unlike policy-based learning approaches that rely on relative or reward-based signals, real offloading decisions depend directly on the correctness of these absolute values. This section formalizes the prediction problem 
and motivates the need for continuous, measurement-driven calibration.

\subsection{Prediction Targets and Context}

Each offloading decision is associated with a context vector $x_t$ describing the execution environment, including server ID, task type, mobility state, and recent load indicators. The goal is to predict absolute values $\hat{y}_t$ 
according to \eqref{eq:absolute_value},
\begin{equation}
    \label{eq:absolute_value}
    \hat{y}_t = (\widehat{\text{RTT}}_t, \widehat{\text{processing}}_t, \widehat{\text{sm\_util}}_t, \widehat{\text{transmission}}_t),
\end{equation}
which directly determine whether a server can meet functional requirements such as latency deadlines or utilization constraints.

In highly dynamic backend systems, these performance metrics exhibit substantial variability due to fluctuating network conditions, heterogeneous hardware, and mobility-induced changes in load and connectivity. 
Their empirical distributions are broad, multimodal, and non-Gaussian, making absolute-value prediction particularly sensitive to drift and distribution shifts. As a result, 
static or purely offline-trained predictors tend to drift over time and cannot maintain reliable accuracy without continuous adaptation.

\subsection{Prediction Error and Decision Correctness}

Let $y_t$ denote the actual measured performance vector. The prediction error is defined according to \eqref{eq:pred_error}.
\begin{equation}
    \label{eq:pred_error}
    e_t = y_t - \hat{y}_t.
\end{equation}
A decision is considered \emph{correct} if the selected server satisfies the functional constraints (e.g., deadline, energy bound) when evaluated with $y_t$, and \emph{incorrect} otherwise. Because performance characteristics 
vary significantly across servers, tasks, and mobility states, even small systematic prediction errors can lead to incorrect decisions and degraded quality of service. This makes absolute-value prediction particularly sensitive 
to drift and distribution shifts.

\subsection{Feedback and the Need for Online Calibration}

After each offloading action, the system observes the performance $y_t$, which serves as feedback. In dynamic backend environments, performance distributions shift over time due to mobility, contention, and resource 
variability. Predictive models therefore degrade unless they are continuously recalibrated using fresh execution measurements.

We formulate the problem as an online prediction task with feedback according to \eqref{eq:problem}:
\begin{equation}
    \label{eq:problem}
    \hat{y}_{t+1} = f_{\theta_t}(x_{t+1}), \qquad
    \theta_{t+1} = \theta_t + \Delta(\hat{y}_t, y_t),
\end{equation}
where $\Delta(\cdot)$ is a lightweight update rule that incorporates execution measurements to correct prediction drift and compensate for underrepresented or newly emerging latency regimes.

This formulation motivates the measurement-driven closed-loop mechanism introduced in this work, which continuously aligns predictive models with real system behavior and thereby improves decision correctness under 
heterogeneous and dynamically shifting execution conditions.

\section{Online-Learning Feedback Loop Design}\label{sec:design}

This work extends the previously introduced four-layer offloading pipeline by integrating a fully operational online-learning feedback loop. The loop continuously refines absolute-value predictions for 
transmission time, processing time, \gls{rtt}, and utilization. It consists of three coordinated components: (i) a data ingestion service, (ii) a feedback processor that aligns predicted and measured 
values, and (iii) a streaming refinement module that performs prediction correction and lightweight online updates.

\subsection{Architectural Overview}

\begin{figure}[t]
    \centering
    \includegraphics[width=\linewidth]{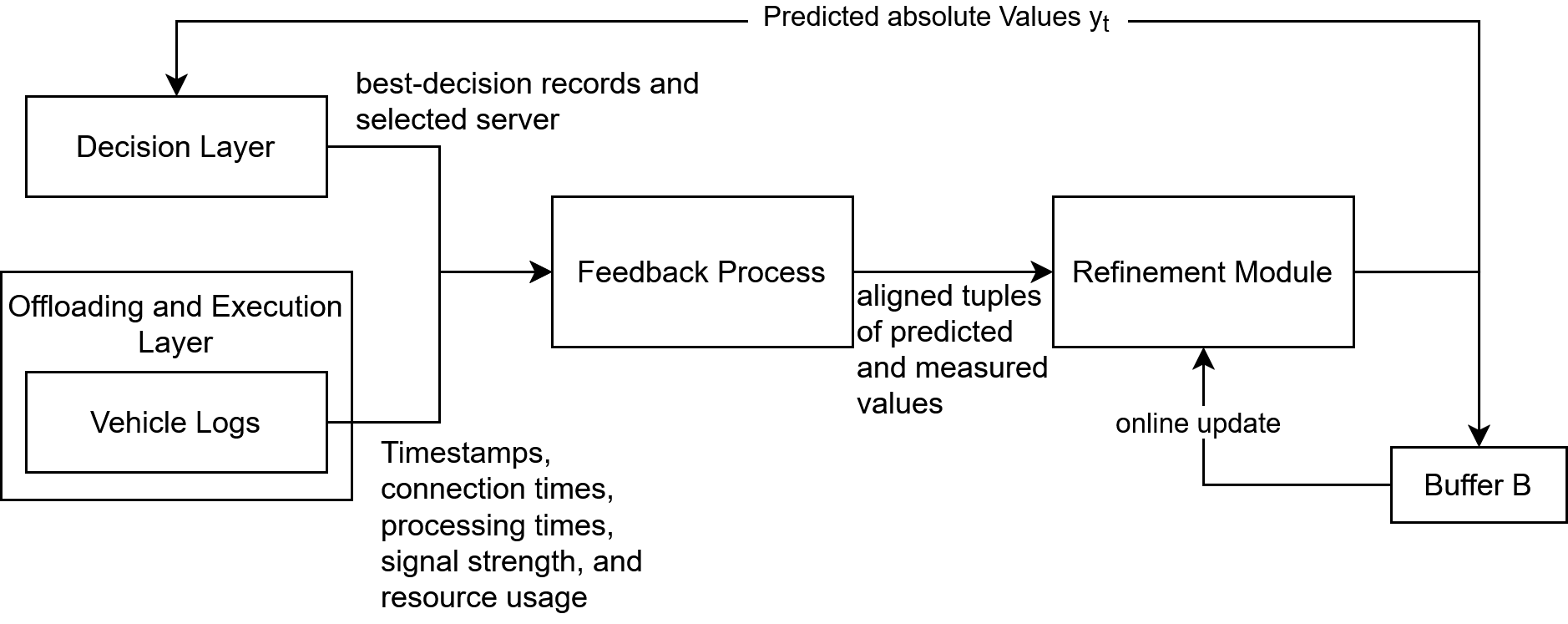}
    \caption{System architecture of the online-learning feedback loop, showing data ingestion, feedback alignment, feature construction, and online refinement.}
    \label{fig:feedback-architecture}
\end{figure}

Figure~\ref{fig:feedback-architecture} illustrates the structure of the feedback loop. The Decision Layer produces predicted metrics and selected servers, while the vehicle logs provide the corresponding measured execution values.
 The feedback processor continuously aligns these streams and forwards matched samples to the refinement module. The refinement module constructs temporal and contextual features, produces refined absolute predictions, and performs 
 periodic online updates once sufficient new samples have accumulated. This closes the loop between observed performance and future predictions.

\subsection{Key Design Decisions}

The design of the feedback loop is guided by four central considerations:

\begin{enumerate}
\item \textbf{Absolute-value prediction:} The system predicts absolute \gls{rtt}, processing time, and utilization rather than relative or reward-based quantities. This preserves compatibility with existing decision algorithms and avoids 
coupling the Decision Layer to a specific learning method.

\item \textbf{Temporal and contextual modeling:} Execution behavior depends on short-term history (e.g., bursts, jitter) and contextual factors such as signal strength, function type, and diurnal patterns. The refinement module 
therefore combines a sliding window of recent measurements with contextual metadata to capture both temporal dynamics and environmental conditions.

\item \textbf{Lightweight multi-head \gls{mlp} architecture:} 
A multi-head \gls{mlp} is used as prediction model because it provides a favorable balance between expressiveness, stability, and online update efficiency. 
\glossary{mlp}s can represent nonlinear relationships between context, temporal features, and performance metrics without requiring recurrent state or complex sequence modeling. 
The shared feature extractor captures general execution patterns, while task-specific heads allow specialization for different function types. 
Compared to heavier architectures such as Recurrent Neural Networks (RNNs) or Transformers, the \gls{mlp} enables fast inference and low-latency online updates, which is essential for real-time offloading decisions and continuous calibration on resource-constrained edge devices.

\item \textbf{Server-wise update isolation:} 
Each backend server maintains its own update buffer and triggers online updates independently. 
This prevents interference between heterogeneous latency regimes and ensures that the model adapts to each server's operating characteristics without cross-contamination.

\item \textbf{Server-specific normalization:} Backend servers operate in fundamentally different latency regimes. Server-specific input and target normalization ensures that the model can represent these heterogeneous ranges without requiring separate models per server.
\end{enumerate}

\subsection{Online Update Strategy}
The refinement module performs online updates once a sufficient number of new matched samples has accumulated. 
Updates are executed \emph{per server} to respect the strong heterogeneity of backend latency regimes and to avoid interference between servers operating in different value ranges. Each server maintains its own update buffer and triggers 
an update once 200 new aligned samples have been collected. This batch size provides a compromise between responsiveness and stability and was empirically chosen as the smallest value that avoids oscillatory behavior while still enabling 
timely adaptation under realistic request rates. It is large enough to smooth out short-term fluctuations, yet small enough to react to regime shifts caused by mobility, contention, or load changes.

To ensure stable adaptation, only the task-specific heads of the multi-head \gls{mlp} are updated online, while the shared feature extractor and all scalers remain fixed. 
Keeping the scalers constant prevents drift in the normalization space and ensures that the model continues to operate within the value ranges learned during offline training. 
This design avoids oscillatory behavior and catastrophic forgetting, which would otherwise occur if the shared representation or normalization layers were updated on small, noisy batches.

Updates are performed using small gradient steps on the task-specific heads, with balanced sampling across task types to prevent bias toward frequently occurring functions. 
This lightweight update rule enables the model to compensate for prediction drift, underrepresented latency regimes, and sudden changes in backend behavior, while preserving the stability required for real-time offloading decisions.

\section{Evaluation} \label{sec:evaluation}

\subsection{Evaluation Setup}

\begin{table}[b]
\centering
\caption{Model and Feedback Loop Parameters}
\label{tab:model-params}
    \begin{tabular}{ll}
        \hline
            \textbf{Parameter} & \textbf{Value} \\
        \hline
            Model type & Multi-head \glossary{mlp} (shared encoder) \\
            Shared layers & FC(128) + ReLU, FC(64) + ReLU \\
            Task-specific heads & FC(64) + ReLU, FC(64) + ReLU, FC(4) \\
            Dropout & 0.1 (shared representation) \\
            Sliding window size $W$ & 15 samples \\
            Input features & $4W$ actuals + 8 rolling + 9 context features \\
            Input dimensionality & $4W + 25$ \\
            Normalization & Server-specific StandardScaler (offline pre-fit) \\
            Prediction smoothing & Rolling average over 3 outputs \\
            Clipping range & $[0, 4500]$ ms \\
            Online update trigger & 200 matched samples \\
            Updated parameters & Task-specific heads only \\
            Optimizer & Adam \\
            Learning rate & $3 \times 10^{-5}$ \\
            Loss weighting & Higher weight for non-critical task head \\
        \hline
    \end{tabular}
\end{table}

The evaluation focuses on the behavior of the online-learning feedback loop under real execution conditions. A simulated vehicle continuously requests the execution of two perception functions (object recognition and emotion recognition), which are deployed 
as containerized microservices on heterogeneous backend hardware. The backend consists of two different Kubernetes clusters located in Mannheim, Germany (two servers; each two \gls{vcpu}, 4 GB \gls{ram} and 60 GB disk capacity) 
and Stuttgart, Germany (four servers; each 8 vCPUs, 16 GB RAM and 125 GB disk capacity). Real execution times, network delays, and resource usage are recorded and fed back into the system, forming a closed-loop offloading environment with naturally emerging variability.

To enable scalable experiments, the physical server locations are mapped to virtual coordinates, while all performance values originate from real executions. The initial prediction model is trained offline for each server separately using execution samples 
collected during initial system operation. During operation, the feedback loop continuously aligns predicted and measured values and refines the model outputs in real time.

Table~\ref{tab:model-params} summarizes the configuration of the multi-head \glossary{mlp} used for prediction refinement. The model processes a sliding window of $W=15$ past samples together with contextual metadata. Online updates are triggered server-wise, once 
200 new matched samples for a specific server are available. Only the task-specific heads are updated, while the shared layers remain frozen to ensure stable adaptation.

\begin{figure}[t]
    \centering
    \includegraphics[width=\linewidth]{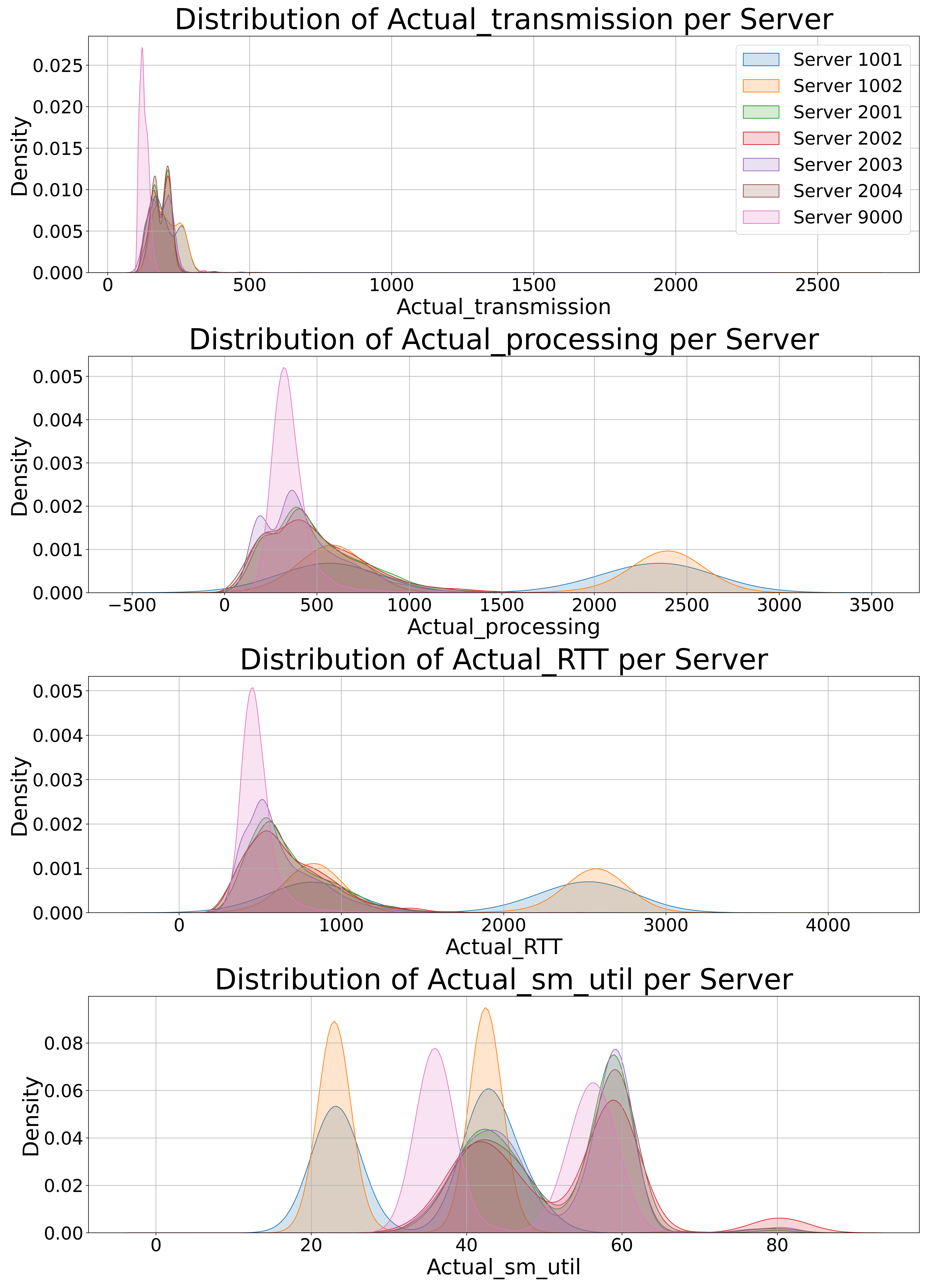}
    \caption{Empirical distributions of the transmission time, processing time, \gls{rtt} and sm\_util across all backend servers. The heterogeneous and multimodal shapes illustrate the variability of real execution conditions for different functions and servers.}
    \label{fig:placeholder}
\end{figure}

The error metrics used in this evaluation follow their standard definitions for \gls{mae}, \gls{mse}, and Mean \gls{mape}. These metrics are computed both as aggregated values per server 
(Table~\ref{tab:error_metrics}) and as incremental curves (Figure \ref{fig:incremental_RTT}) over time to analyze the effect of online learning. In addition, a sigma-based error analysis is performed 
to characterize the variability of the prediction errors and to complement the absolute error metrics (Table~\ref{tab:sigma_analysis_from_csv}).

\begin{figure*}[tb]
    \centering
    \includegraphics[width=\linewidth]{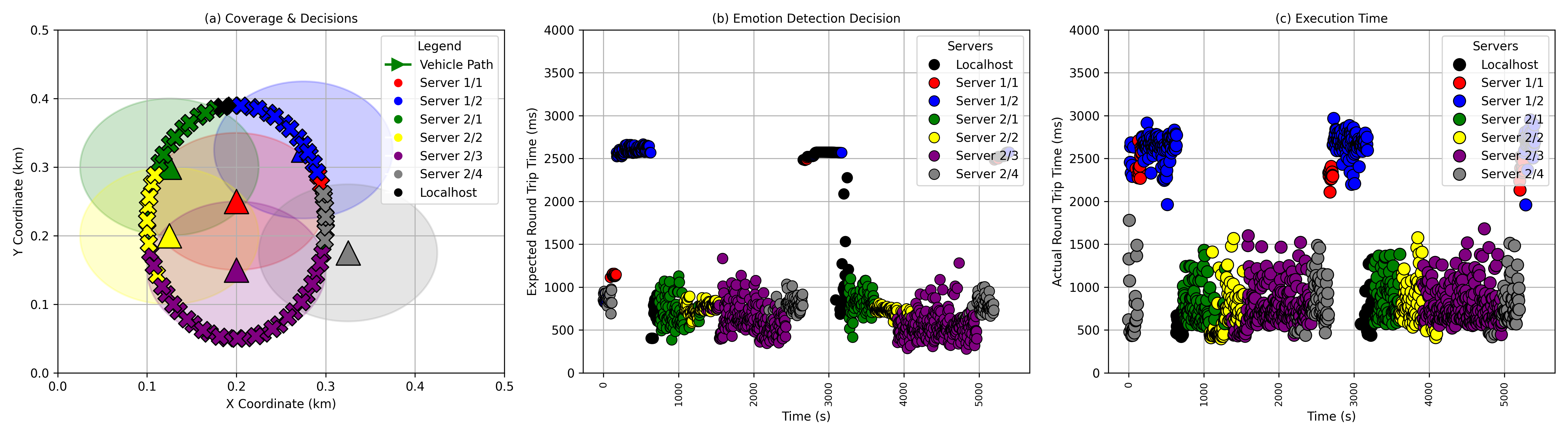}
   \caption{Offloading behavior for the emotion-recognition task. 
    Subplot~(a) shows the vehicle trajectory, server positions, and resulting offloading decisions. 
    Subplot~(b) depicts the predicted RTTs of the multi-head \glossary{mlp}, including fluctuations from heterogeneous server performance and online updates. 
    Subplot~(c) shows the measured RTTs, which exhibit a wider and more irregular latency distribution while preserving the overall performance trends.}
    \label{fig:coverage_decisions}
\end{figure*}

\subsection{Characteristics of the Input Data}\label{sec:characteristic}
The feedback loop operates on real execution measurements collected from all backend servers during system operation. These measurements exhibit substantial heterogeneity across servers and tasks, reflecting differences in hardware capabilities, network paths, 
and runtime load conditions. Figure~\ref{fig:placeholder} summarizes the empirical distributions of the four target metrics, \textit{processing time}, \textit{\gls{rtt}}, \textit{system (sm)\_util}, and \textit{transmission time}, across all servers.

Across servers, all metrics show broad, non-Gaussian, and often multimodal distributions, indicating multiple operating regimes such as idle versus loaded states or stable versus congested network conditions. The variance of these metrics is consistently high, 
with several servers exhibiting long-tailed or sharply peaked behaviors. Processing times differ markedly between lightweight and compute-intensive services and servers, \gls{rtt} ranges from tight low-latency clusters to wide high-variance regimes, SM utilization forms 
distinct computation complexity of a service and scheduling-related bands, and transmission times vary sharply based on backend data center connection.

This combination of high variance and cross-server heterogeneity makes absolute-value prediction inherently difficult: the same task may experience fundamentally different performance characteristics depending on the selected server. As a result, even small 
prediction errors can have asymmetric effects, from negligible impact on lightly loaded servers to deadline violations on congested ones.

\subsection{Evaluation Results}
We evaluate how spatial coverage, predicted latency, and measured execution times influence the offloading behavior of the closed-loop pipeline. All real backend servers from both clusters, as well as local execution, are considered. Offloading requests are 
generated continuously along the vehicle trajectory while the offloading decision is continuously made based on predicted and measured RTTs. The measured values vary substantially over time (see also Section~\ref{sec:characteristic}). Figure~\ref{fig:coverage_decisions} 
summarizes the resulting behavior across three subplots using consistent color coding.

Subplot~(a) shows the simulated vehicle trajectory, server positions, communication ranges, and the corresponding offloading decisions. The vehicle executes two perception functions every 5\,s; for clarity, only every 40th decision for emotion recognition is shown as \textit{X} marker. 
As the vehicle moves through overlapping coverage regions, the set of feasible servers changes dynamically. Decisions are assigned to the server with the lowest expected latency within range, with local execution used when no server is reachable or predicted values 
are unfavorable.

Subplot~(b) presents the offloading decisions derived from the predicted RTTs of the multi-head \glossary{mlp} model. Although only a single function is shown, the predicted RTTs vary substantially due to heterogeneous server performance, resulting in distinct latency regimes. 
All prediction samples are displayed, revealing the full temporal variability. Online-learning updates occasionally shift the predictions, producing visible changes in the decision pattern. As the vehicle transitions between coverage zones, these fluctuations lead 
to alternating server selections, with local execution chosen when predicted RTTs exceed acceptable bounds. The subplot represents a cold start scenario.

\begin{table}[b]
    \centering
    \caption{Comparison of absolute error metrics for a well-performing server (2003) and a poorly-performing server (1001).}
        \begin{tabular}{l l r r r}
            \toprule
                Server & Target & \gls{mae} & \gls{mse} & \gls{mape} \\
                \midrule
                    2003 & Transmission [ms] & 22.87 & 1384.98 & 13.72 \\
                    2003 & Processing [ms]  & 185.66 & 65188.56 & 49.56 \\
                    2003 & \gls{rtt} [ms]         & 188.93 & 63352.94 & 31.63 \\
                    2003 & Util [\%]        & 7.48 & 140.99 & 13.38 \\
                \midrule
                    1001 & Transmission [ms] & 77.63 & 12548.45 & 42.23 \\
                    1001 & Processing [ms]  & 621.90 & 722304.42 & 33.55 \\
                    1001 & \gls{rtt} [ms]        & 568.26 & 531390.88 & 27.85 \\
                    1001 & Util [\%]        & 15.88 & 515.04 & 33.97 \\
            \bottomrule
        \end{tabular}
    \label{tab:error_metrics}
\end{table}

\begin{figure*}[t]
    \centering
    \includegraphics[width=\linewidth]{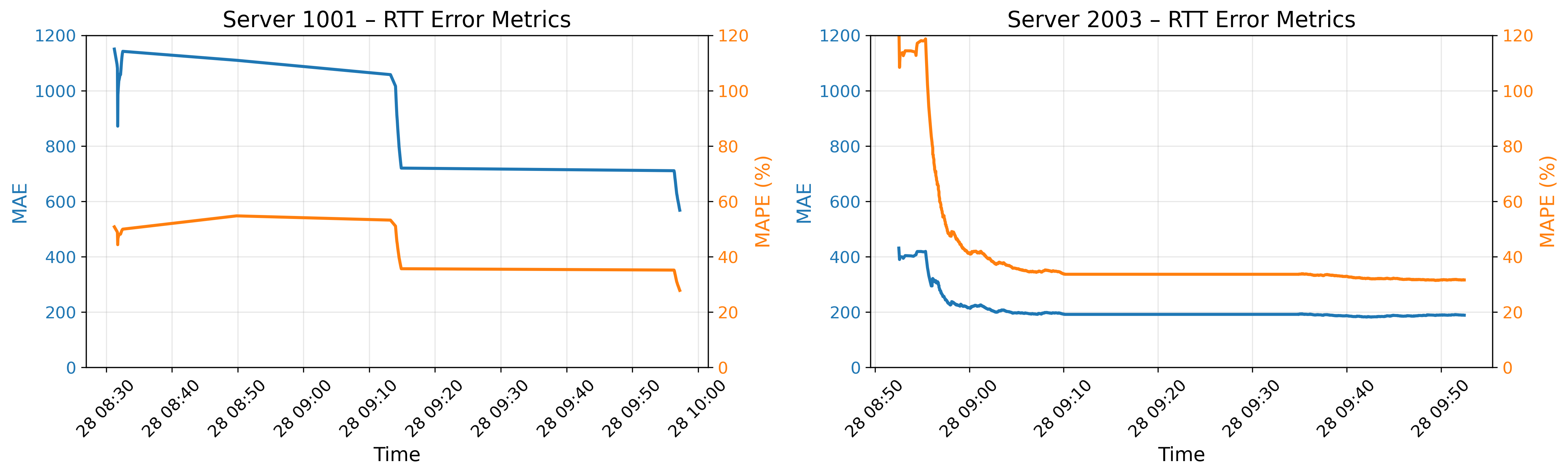}
    \caption{Incremental \gls{mae} and \gls{mape} for \gls{rtt} on Server~1001 and Server~2003. Both curves exhibit the characteristic step-like reductions caused by online-learning updates. Server~2003 converges faster and stabilizes at lower error levels due to its more 
    consistent execution behavior and higher sample rate, whereas Server~1001 shows larger update steps and higher residual errors, reflecting its higher \gls{rtt} variability and lower request frequency.}

    \label{fig:incremental_RTT}
\end{figure*}

Subplot~(c) shows the actual RTTs measured on the Kubernetes backend. Compared to the predicted values in Subplot~(b), the real system exhibits a much wider and more irregular latency distribution with pronounced variance and occasional extremes. Although the 
online-learning mechanism continuously updates the task-specific heads, the model still underestimates the full range of \gls{rtt} fluctuations, indicating that the heads were not sufficiently trained to capture the broader latency regimes. Nevertheless, the model 
preserves the general trend and relative ordering of server performance. This subplot therefore highlights the remaining gap between model-based predictions and real-world variability, even under active online adaptation.

Table~\ref{tab:error_metrics} contrasts one of the best-performing servers (2003) with one of the worst-performing servers (1001). Server~2003 exhibits consistently low transmission errors and moderate processing and \gls{rtt} deviations, indicating stable execution
characteristics and relatively homogeneous latency behavior. In contrast, Server~1001 shows markedly higher \gls{mae} and \gls{mse} values across all targets, particularly for processing time and \gls{rtt}, reflecting strong variability and frequent latency spikes in the 
underlying hardware and network path.

The incremental error plots in Figure~\ref{fig:incremental_RTT} further illustrate these differences. Both servers show the characteristic step-like reductions in \gls{mae} and \gls{mape} caused by online-learning updates, demonstrating that the feedback loop 
continuously incorporates new execution measurements. Server~2003 produces smaller and faster-converging steps due to its stable behavior and higher sample rate. Server~1001, in contrast, receives fewer requests and operates at substantially higher \gls{rtt} levels, 
which leads to larger absolute errors and more pronounced update steps. Despite the corrective updates, residual errors on Server~1001 remain significantly higher, indicating that the task-specific heads cannot fully capture the server's highly variable performance regime.

The update-effect analysis quantifies these observations. For Server~1001, the \gls{rtt}-\gls{mae} decreases from 1150.6\,ms to 568.3\,ms after the first online update, corresponding to a 50.6\% reduction. Server~2003 shows a similar trend, reducing its \gls{rtt}-\gls{mae} from 
431.4\,ms to 182.3\,ms (57.8\% reduction). These results highlight that the feedback loop is essential for compensating fluctuating or underrepresented operating regimes, enabling the model to adapt both to highly variable servers and to comparatively stable ones.

The sigma-based error analysis in Table~\ref{tab:sigma_analysis_from_csv} provides a detailed characterization of the residual distributions for the processing, \gls{rtt}, and transmission metrics. For each server-target pair, the table reports 
the mean bias, the empirical residual standard deviation (Sigma), and the percentage of samples whose absolute errors exceed one, two, or three standard deviations. In addition, the table lists the share of measurements whose absolute deviation 
surpasses the operational tolerance threshold of 100\,ms. The tolerance threshold of 100\,ms reflects a conservative bound for latency-sensitive perception and decision functions, beyond which deviations are likely to affect end-to-end timing guarantees.
The Util target is not considered here. 

Overall, the results reveal clear differences across servers. The servers in the 2000-series exhibit comparatively low mean biases and moderate Sigma values, indicating more stable execution characteristics and narrower latency distributions 
than the 1000-series. While their proportions of samples exceeding 1$\sigma$, 2$\sigma$, and 3$\sigma$ remain moderate, the absolute tolerance violations above 100\,ms are still substantial for Processing and \gls{rtt}. In contrast, 
the 1000-series servers, most notably Server~1001, show markedly larger residual spreads, higher Sigma values, and an even greater fraction of samples beyond the 100\,ms threshold, reflecting their heterogeneous and highly variable performance regimes.

\begin{table*}[h]
\centering
\caption{Sigma-Based Error Analysis per Server and Target.}
\begin{tabular}{llrrrrrr}
\hline
Server & Target & Mean-Bias & Sigma & $>1\sigma$ [\%] & $>2\sigma$ [\%] & $>3\sigma$ [\%] & $>\text{Tol} [\%]$ \\
\hline
1001 & Processing    & -246.14 & 524.25 & 33.12 & 7.69 & 0.64 & 77.56 \\
1001 & \gls{rtt}     & -170.49 & 466.94 & 34.83 & 4.91 & 0.64 & 78.84 \\
1001 & Transmission  & 95.60   & 99.90  & 61.32 & 12.82 & 0.43 & 61.32 \\
1001 & Util          & 11.97   & 9.70   & 53.63 & 26.07 & 5.77 & -- \\
\hline
1002 & Processing    & 49.26   & 171.74 & 13.01 & 1.63 & 0.80 & 32.37 \\
1002 & \gls{rtt}     & 14.95   & 182.91 & 9.83  & 2.28 & 1.00 & 27.28 \\
1002 & Transmission  & 8.73    & 70.93  & 2.23  & 1.38 & 0.73 & 1.58 \\
1002 & Util          & -2.90   & 5.02   & 19.66 & 4.58 & 3.80 & -- \\
\hline
2001 & Processing    & -3.49   & 236.73 & 27.12 & 4.52 & 1.07 & 72.82 \\
2001 & \gls{rtt}     & -38.61  & 238.85 & 33.11 & 3.83 & 0.88 & 74.95 \\
2001 & Transmission  & -9.15   & 28.04  & 23.01 & 2.48 & 1.00 & 0.97 \\
2001 & Util          & 0.48    & 4.93   & 16.26 & 3.39 & 1.57 & -- \\
\hline
2002 & Processing    & 34.40   & 252.69 & 24.51 & 5.21 & 1.84 & 72.11 \\
2002 & \gls{rtt}     & -6.83   & 255.92 & 26.35 & 4.19 & 1.23 & 76.81 \\
2002 & Transmission  & -11.55  & 35.80  & 17.06 & 2.45 & 2.45 & 2.45 \\
2002 & Util          & -0.35   & 10.30  & 7.97  & 2.25 & 1.74 & -- \\
\hline
2003 & Processing    & 92.68   & 214.86 & 25.02 & 8.55 & 2.21 & 49.56 \\
2003 & \gls{rtt}     & 46.85   & 226.45 & 22.05 & 6.12 & 1.53 & 60.48 \\
2003 & Transmission  & -12.72  & 78.41  & 1.44  & 0.99 & 0.39 & 1.29 \\
2003 & Util          & -3.30   & 5.33   & 38.30 & 3.63 & 0.88 & -- \\
\hline
2004 & Processing    & 46.40   & 233.05 & 25.42 & 5.92 & 1.50 & 65.56 \\
2004 & \gls{rtt}     & -1.87   & 235.12 & 25.69 & 5.09 & 1.27 & 72.59 \\
2004 & Transmission  & -7.43   & 30.03  & 18.44 & 2.49 & 1.44 & 1.43 \\
2004 & Util          & -2.63   & 6.92   & 20.82 & 1.77 & 0.94 & -- \\

\hline
\end{tabular}
\label{tab:sigma_analysis_from_csv}
\end{table*}

\section{Discussion} \label{sec:discussion}

The evaluation shows that the proposed measurement-driven feedback loop can maintain stable prediction quality under real execution conditions, despite the substantial heterogeneity and temporal variability of the backend environment. Rather than eliminating these 
effects, the feedback loop enables the prediction layer to continuously track them, which is essential for reliable absolute-value prediction in dynamic edge systems.

A first key insight is the pronounced heterogeneity across backend servers. As the empirical distributions and server-wise error metrics demonstrate, servers operate in fundamentally different latency regimes driven by hardware class, network path, and instantaneous 
load. These differences are not artifacts of the model but intrinsic properties of the physical system. The feedback loop therefore serves as a mechanism to follow these evolving regimes rather than to homogenize them.

A second insight concerns the gap between predicted and measured RTTs. The real system exhibits long-tailed, multimodal, and highly irregular latency distributions that exceed the representational capacity of the lightweight task-specific heads. This is an 
intentional design trade-off: the model prioritizes responsiveness, fast online updates, and stability over full distribution modeling. The higher absolute errors observed on servers such as 1001 stem not only from their volatile behavior but also from the 
substantially larger value ranges the model must learn. These effects highlight the inherent difficulty of absolute-value prediction in heterogeneous edge environments.

A third insight concerns the role of the feedback loop itself. The update-effect analysis shows that online calibration is essential for compensating fluctuating or underrepresented operating regimes. 
Servers such as 1001 benefit from large update-induced error reductions due to their volatile behavior, whereas more stable servers converge quickly with smaller corrections.

Taken together, the update-effect analysis and the sigma-based error characterization provide a coherent view of the role of the feedback loop in a heterogeneous backend environment. The feedback mechanism does not reduce the inherent variability of the servers, nor 
is this its objective, but instead mitigates systematic drift by continuously incorporating new execution measurements. The sigma-based metrics show that substantial deviations and tolerance violations persist, particularly on highly volatile servers, reflecting the 
underlying and unavoidable variability of real-world latency regimes. At the same time, the update-effect analysis demonstrates that the online updates consistently reduce bias and improve alignment with the current operating conditions. Overall, the feedback loop 
stabilizes the prediction process by counteracting drift and maintaining model alignment, even though large residual fluctuations remain an intrinsic property of the backend systems.

Two practical limitations arise from the proposed online-learning setup. First, updates require batches of a relevant number of matched samples, which accumulate at different rates depending on mobility and coverage. This explains the extended plateaus in the incremental 
error curves. Second, the sliding window introduces a cold-start phase in which the model lacks sufficient temporal context, leading to temporarily unstable predictions until enough history is available.

Overall, the results demonstrate that the feedback loop effectively counteracts prediction drift and preserves the correct relative ordering of backend servers, even under strong variability. The remaining discrepancies between predicted and measured values 
reflect structural trade-offs between model complexity, update frequency, and the heterogeneous data distribution rather than shortcomings of the approach. The presented mechanism provides a practical foundation for future extensions such as uncertainty-aware 
prediction, adaptive update triggers, or more advanced hybrid models that combine lightweight online refinement with periodic offline retraining.

\section{Conclusion} \label{sec:conclusion}

This work presents a fully operational, measurement-driven feedback loop for computation offloading pipelines, enabling continuous refinement of absolute-value predictions under real execution conditions. By aligning predicted and measured metrics at 
runtime and updating the model incrementally, the system closes a link between prediction and actual backend behavior. The evaluation demonstrates that this closed-loop design counteracts prediction drift and maintains functional prediction quality 
despite substantial variability in network conditions, hardware characteristics, and temporal load dynamics.

The results also highlight structural challenges inherent to absolute-value prediction in heterogeneous edge environments. The empirical data exhibit broad, multimodal, and server-specific latency regimes, which fundamentally limit the achievable 
accuracy of lightweight models. These discrepancies reflect expected trade-offs between model complexity, update frequency, and the need for stable incremental learning rather than shortcomings of the approach. The following key takeaways emerge:

\begin{itemize}
    \item A lightweight closed-loop mechanism can maintain usable prediction quality under real-world variability and counteract prediction drift in dynamic backend environments.
    \item Strong cross-server heterogeneity and multimodal latency distributions make absolute-value prediction inherently difficult, underscoring the need for continuous measurement-driven calibration.
    \item Practical constraints such as batch-based updates and cold-start behavior shape the temporal dynamics of online learning and motivate more adaptive update strategies.
\end{itemize}

In summary, the proposed feedback loop demonstrates that continuous measurement-driven refinement is not merely beneficial but necessary for maintaining robust absolute-value predictions in heterogeneous edge environments. Without online calibration, 
prediction drift and underrepresented latency regimes would quickly degrade decision quality. While the current design prioritizes stability and responsiveness over full distribution modeling, it provides a solid foundation for future extensions, 
including uncertainty-aware prediction, adaptive windowing, and more flexible online update mechanisms to further improve robustness under highly variable execution conditions.

\section*{Acknowledgment}
The authors disclose that generative AI has been used for improving the grammar and language of the paper. The authors have reviewed and edited all content as needed and take full responsibility for the scientific integrity and authenticity of this article.

\bibliographystyle{IEEEtran}
\bibliography{bib}

\end{document}